\documentclass{osa-article}

\usepackage{amsmath}
\usepackage{graphicx}
\usepackage{svg}
\usepackage{verbatim}
\journal{josab}
\articletype{Research Article}

\begin{document}

\title{Design considerations of photonic lanterns for diffraction-limited spectrometry}

\author{
Jonathan Lin,\authormark{1,*} 
Nemanja Jovanovic,\authormark{2} and
Michael P. Fitzgerald\authormark{1}
}

\address{\authormark{1} Physics \& Astronomy Department, University of California, Los Angeles (UCLA), 475 Portola Plaza, Los Angeles 90095, USA}
\address{\authormark{2}Department of Astronomy, California Institute of Technology, 1200 E. California Blvd., Pasadena, CA 91125, USA}

\email{\authormark{*}jon880@astro.ucla.edu}

\begin{abstract}
The coupling of large telescopes to astronomical instruments has historically been challenging due to the tension between instrument throughput and stability. Light from the telescope can either be injected wholesale into the instrument, maintaining high throughput at the cost of point-spread function (PSF) stability, or the time-varying components of the light can be filtered out with single-mode fibers (SMFs), maintaining instrument stability at the cost of light loss. Today, the field of astrophotonics provides a potential resolution to the throughput-stability tension in the form of the photonic lantern (PL): a tapered waveguide which can couple a time-varying and aberrated PSF into multiple diffraction-limited beams at an efficiency that greatly surpasses direct SMF injection. As a result, lantern-fed instruments retain the stability of SMF-fed instruments while increasing their throughput. To this end, we present a series of numerical simulations characterizing PL performance as a function of lantern geometry, wavelength, and wavefront error (WFE), aimed at guiding the design of future diffraction-limited spectrometers. These characterizations include a first look at the interaction between PLs and phase-induced amplitude apodization (PIAA) optics. We find that Gaussian-mapping beam-shaping optics can enhance coupling into 3-port lanterns but offer diminishing gains with larger lanterns. In the $y$- and $J$-band (0.97--1.35 $\upmu$m) region, with moderately high WFE ($\sim$10\% Strehl ratio), a 3-port lantern in conjunction with beam-shaping optics strikes a good balance between pixel count and throughput gains. If pixels are not a constraint, and the flux in each port will be dominated by photon noise, then larger port count lanterns will provide further coupling gains due to a greater resilience to tip-tilt errors. Finally, we show that lanterns can maintain high operating efficiencies over large wavelength bands where the number of guided modes at the lantern entrance drops, if care is taken to minimize the attenuation of weakly radiative input modes. 
\end{abstract}

\section{Introduction}
The stability of a ground-based spectrometer is primarily limited by atmospheric turbulence, which distorts incoming wavefronts and makes the point-spread function (PSF) of an optical system temporally unstable. These instabilities can be compounded when aberrated light is transferred to an optical fiber if the core diameter of the fiber is large compared to the wavelength of the light. Since large-diameter fibers let light propagate in multiple eigenmodes (specifically, the linearly polarized or LP modes, which apply for weakly guiding and radially symmetric waveguides \cite{fiber}), the power travelling down these ``multi-mode fibers'' (MMFs) will continually shift between the available modes over the course of propagation, due to bends, imperfections, stresses, and strains along the fiber. This is termed modal noise \cite{Baudrand:01,Jovanovic:16}.
While the issue of modal noise has been partially circumvented in the past by scrambling the propagating light with fiber agitators~\cite{plavchan2013-PNI}, in many cases spectrometers fed with single-mode fibers (SMFs) are becoming the more attractive option. 
SMFs have only a single mode of propagation (LP$_{01}$) which automatically eliminates modal noise and provides a static beam profile for the spectrometer even in the presence of a temporally fluctuating input field distribution, greatly improving instrument precision.
Diffraction-limited SMF-fed spectrometers are also significantly more compact than traditional seeing-limited spectrometers, reducing cost and improving stability against thermal and mechanical deflections~\cite{BlandHawthorn2010-PIMMS,jovanovic2016-ESS}. These gains come at the expense of instrument throughput. Because the LP$_{01}$ mode is flat in phase and Gaussian-like in amplitude, efficient coupling into SMFs is difficult at ground-based observatories, resulting in greater light losses in SMF-based fiber injectors over MMF-based fiber injectors.
\begin{figure}
    \centering
    \includegraphics[width=\textwidth]{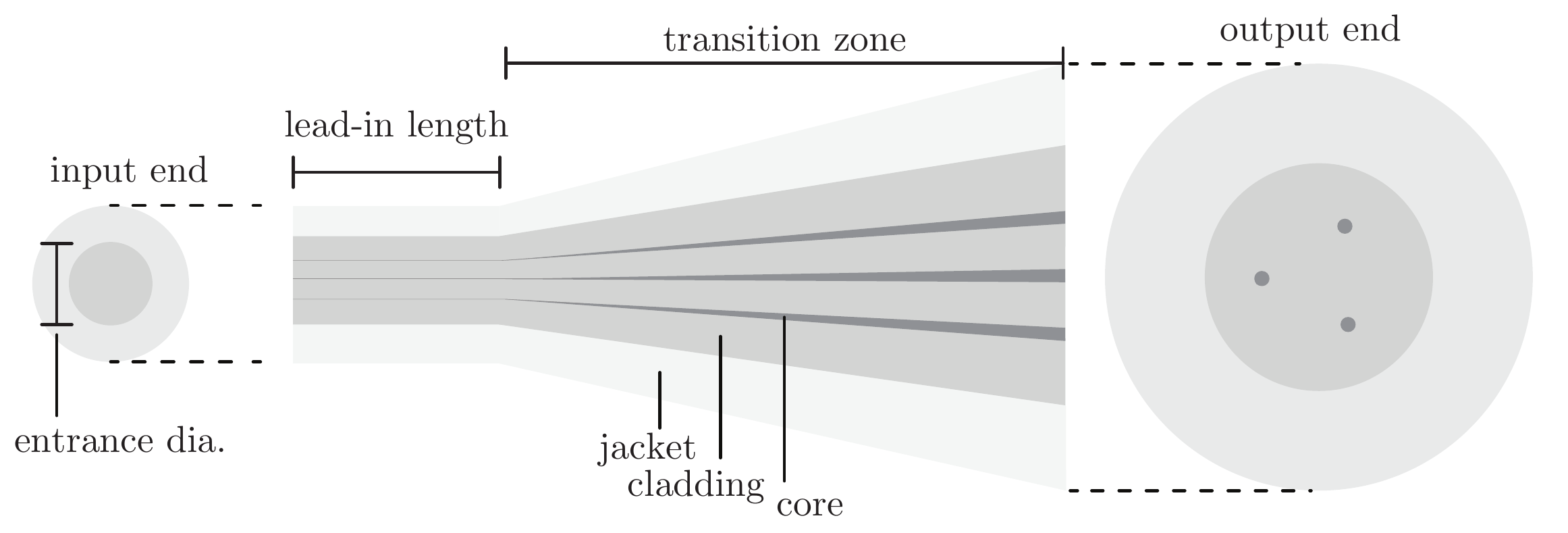}
    \caption{The photonic lantern, a tapered waveguide that can adiabatically transfer light from a multi-mode fiber-like input end into multiple single-mode cores. In particular, the lanterns considered in this work have output geometries similar to multi-core fibers. We choose this architecture for simplicity, and neglect propagation in the single-mode cores after the lantern transition. For the purpose of this work we additionally assume a linear tapering profile between the ``lead-in'' portion of the waveguide and the lantern's output, along with a core-cladding index contrast of $8.8\times 10^{-3}$ and a cladding-jacket contrast of $5.5\times 10^{-3}$. Important geometrical considerations in lantern design include the lead-in length and taper factor (the degree by which SMF cores shrink from output to input).}
    \label{fig:lant}
\end{figure}
\\\\
There are several pathways to increasing throughput while maintaining instrument stability. In SMF coupling, the first hurdle to overcome is the illumination mismatch between the Airy pattern of the telescope beam and the Gaussian-like distribution of the fiber mode. One method to overcome this is by using phase-induced amplitude apodization (PIAA) optics \cite{Guyon:03}. PIAA optics use a pair of lenses (or mirrors) in collimated space to reshape the flat-topped aperture illumination profile of incoming waves into a distribution more amenable for applications such as fiber injection and coronagraphy. In the context of SMF injection, incoming waves are typically reshaped to be Gaussian in amplitude, thereby preventing the formation of diffraction rings in the focal plane. This, in turn, promotes a PSF that better matches the fundamental mode of an SMF, increasing coupling efficiency. Such ``beam-shaping" optics have been shown to work well (e.g. \cite{Jovanovic:17,Calvin:21}, on SCExAO and KPIC respectively). However, PIAA optics only correct for errors in field amplitude, not phase. Phase errors (formed from both atmospheric and instrumental effects) are the second hurdle in efficient SMF coupling. Such errors can be corrected with adaptive optics (AO), but these systems are less effective at shorter wavelengths. For example, at the short end of the near-infrared (NIR) $y$ and $J$ bands (0.97--1.35 $\upmu$m), where the Strehl ratio (the ratio between the peak intensity of an aberrated PSF and the peak intensity of an unaberrated PSF) is low ($<50\%$), residual phase errors will limit the coupling to $<50\%$. These wavelength regimes will be the focus for the next push in direct exoplanet spectroscopy with upcoming instruments such as HISPEC and MODHIS~\cite{mawet2019-HISPEC}.
\begin{figure}
    \centering
    \includegraphics[width=\textwidth]{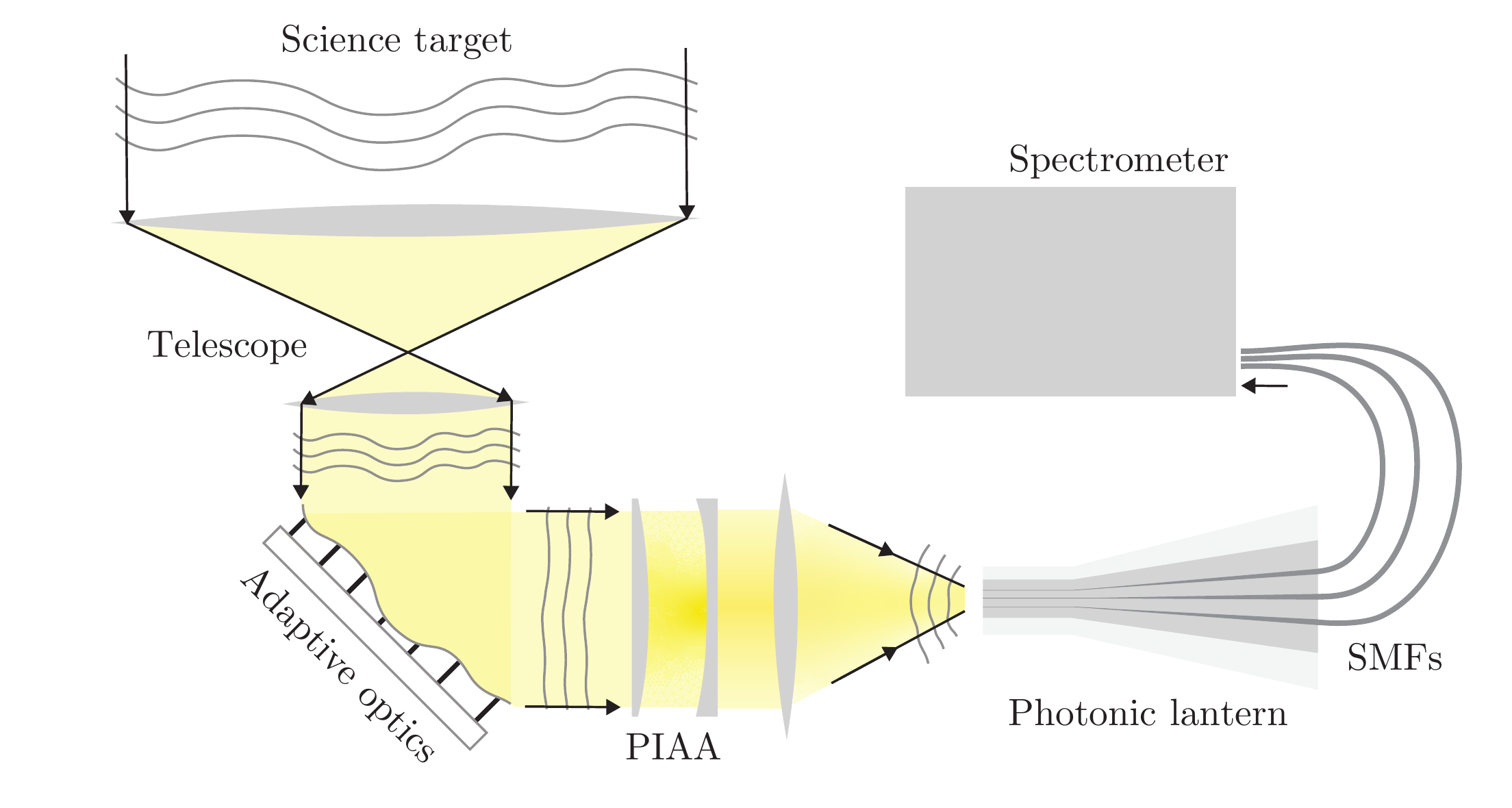}
    \caption{A diagram of a fiber-fed diffraction-limited spectrometer coupled to a telescope with an AO system and a PL fiber injector. The amplitude-remapping effect of PIAA optics is shown through gradients, in yellow.}
    \label{fig:PLinjector}
\end{figure}
An alternative approach is to initially couple aberrated telescope light into a MMF, which can more efficiently accept light from the complex speckle patterns characteristic of low Strehl beams, and ``stabilize" the light before injection into the instrument. The field of astrophotonics makes such a pathway possible via the photonic lantern (PL) \cite{Leon-Saval:14,Birks:15}. The PL is a waveguide that transitions adiabatically from an MMF-like geometry to a multi-core fiber (MCF) geometry, allowing light to move between the MMF-like and MCF-like ends with low loss. Such devices can be formed by heating and tapering one end of either an MCF or a bundle of SMFs (though methods such as laser inscription can also be used \cite{Thomson:11}, for a review of construction methods see \cite{Birks:15}). A diagram of a PL formed from a tapered MCF is shown in Fig.~\ref{fig:lant}. Such devices are already in use with fiber Bragg gratings (e.g. GNOSIS~\cite{Bland-Hawthorn2011}, PRAXIS~\cite{ellis2018a}) and are currently being considered for use in focal plane wavefront sensing \cite{Corrigan:18,Norris:20}. Figure \ref{fig:PLinjector} shows how a PL might be used in the context of high-resolution spectrometry. AO-corrected (and potentially beam-shaped) telescope light is injected into the MMF-like end of a PL; over the course of propagation through the lantern, this light becomes efficiently confined within the single-mode lantern cores, which can then be routed into the science instrument through SMFs. As such, lantern-based fiber injectors retain both the gain in input coupling efficiency offered by MMF-based injectors as well as the gain in instrument stability offered by SMF-based injectors. 
\\\\
In recent years, significant headway has been made in our understanding of PLs and their potential uses in astronomy \cite{Harris:15,Cvetojevic:17,Anagnos:18,Corrigan:18}. PLs are also currently being considered for a number of applications outside astronomy, including mode-selective fiber lasers \cite{Wang:18} and amplifiers \cite{Montoya:17}, spatial mode control in optical fibers \cite{Montoya:16}, LIDAR systems \cite{Ozdur:15}, optical ground receivers for laser satellite communications \cite{Tedder:19}, and free space optical communications \cite{Ozdur:15}. The latter three investigate how PLs can couple light through the atmosphere, similar to what we consider in astronomy.
This work aims at extending existing studies of PLs to the realm of high-resolution spectrometry. In particular, we focus on characterizing the chromatic behavior of PLs, their coupling as function of Strehl, and their performance in the few-port regime (especially relevant for high-resolution spectrometry which requires many pixels): all properties that must be better understood before lanterns can make their way into the next generation of spectrometers. To this end, we present a numerical model for PL performance, as outlined in \S\ref{sec:method}. We then apply this model in \S\ref{sec:res} to showcase the potential gains offered by PLs across a range of geometries, wavelengths, and wavefront errors (WFEs), with the aim of informing the next generation of instrument designs. In the process, we include a first look at the interaction between PLs and beam-shaping PIAA optics. Results are discussed in \S\ref{sec:disc}.

\section{Methodology}\label{sec:method}
On-sky testing with an experimental setup is the ultimate way to determine the true performance of a lantern-based injector. However, before investing in the development of complex infrastructure, it is important to understand the expected performance with a detailed model; such models will also help interpret future experimental results. We present a model for lantern performance based on the statistical assessment of the scalar electric fields at the lantern's input and output. From these fields, we compute our prioritized metric for lantern performance: the lantern throughput, defined as the ratio between total power in the single-mode cores at the lantern's output and the total incident power at the lantern's input (though other metrics, such as the degree of mutual coherence between the lantern output cores, may also be applicable in other contexts). The electric field at the lantern's input is a result of partially compensated aberrations as well as potential beam shaping, while the output electric field depends on the propagation through the lantern device. 
\begin{figure}
    \centering
    \includegraphics[width=\textwidth]{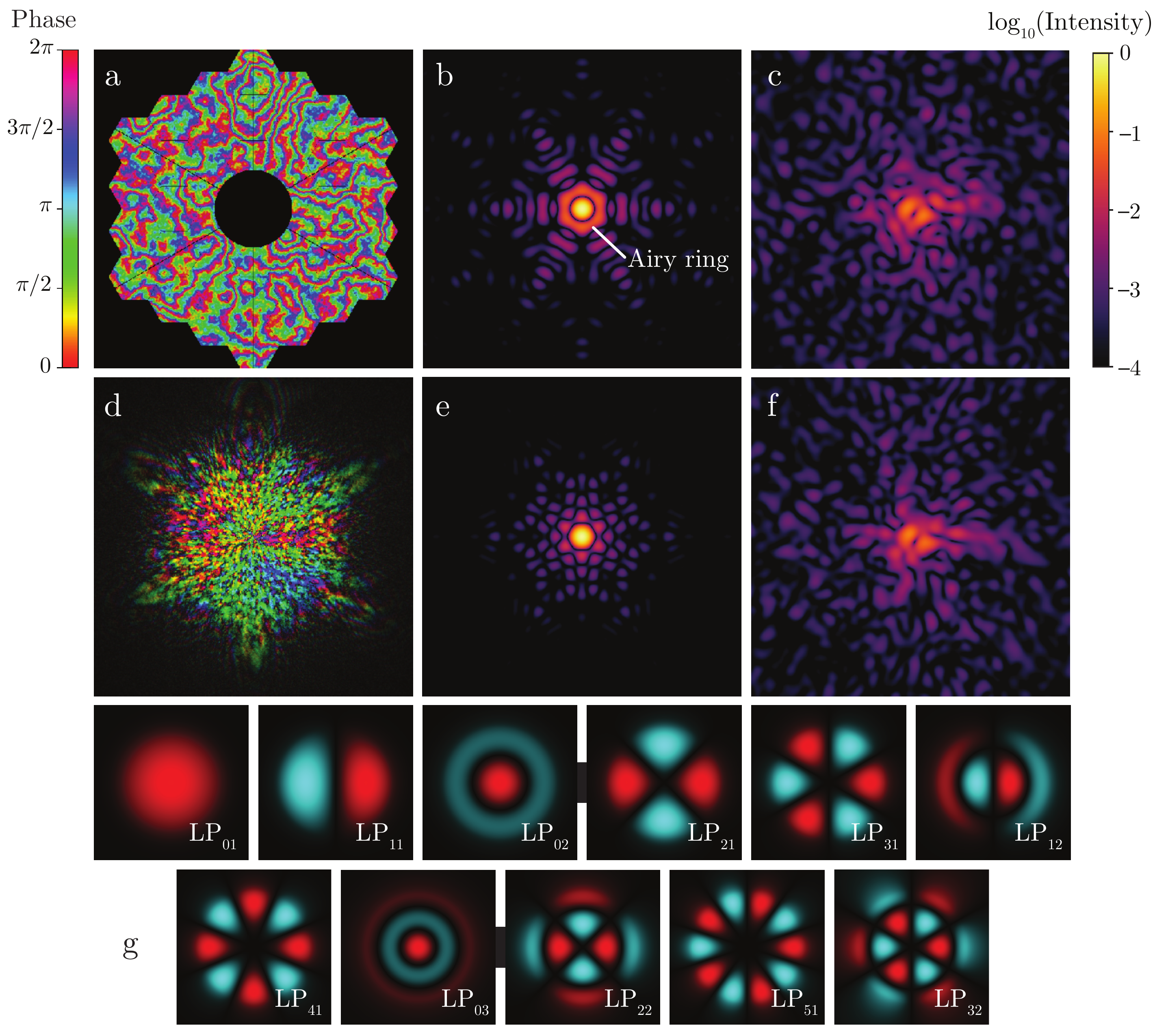}
    \caption{a: the Keck pupil, with turbulence-induced phase errors plotted in color. b: the PSF in the absence of WFE. c: same as b but with WFE ($\sim$10\% Strehl). d: same thing as a, after beam-shaping to a Gaussian illumination profile. e: same thing as b, but with beam-shaping. Note the disappearance of the Airy ring. f: Same as c, but with beam-shaping. g: The first 19 LP$_{lm}$ (linearly polarized) modes, the basis for propagation down a circular, weakly-guiding step-index fiber. Modes are presented in the order they appear as core diameter scales up. Note that all modes with $l\neq 0$ can be rotated by 90$^\circ$ to yield another linearly independent mode; these rotations are not shown. Modes that appear simultaneously are shown connected. Panels a, d, and g display phase information according to the left color bar while panels b, c, e, and f display intensity information according to the right color bar.}
    \label{fig:wfs}
\end{figure}
\\\\
There are multiple approaches to computing lantern throughput. The electric field distribution at the lantern output after beam propagation can be readily converted to a lantern throughput, via overlap integrals (note that this is different than taking the total power of the output field, which may include extra power in the cladding, not guided by the SMF cores). However, under certain circumstances, throughput can also be computed from a ``modal analysis" perspective through careful comparison between the input field and the propagating modes of the lantern. We take both approaches in this work. In this section, we detail our models for WFE (\S\ref{sec:wfe}), AO correction (\S \ref{sec:ao}), beam shaping (\S\ref{sec:piaa_method}), and injection into and propagation through a PL (\S\ref{sec:lant_method}).

\subsection{Wavefront error}\label{sec:wfe}
To compute a coupling efficiency for any sort of fiber injector, we first need a statistical model describing the WFE of light entering our optical system. To this end we assume the standard Kolmogorov turbulence model for a single thin layer of frozen-flow turbulence translating at a velocity of 10 m/s. Realizations of turbulent phase screens are generated over a period of 10 seconds, corresponding to 10 pupil crossings over the course of a simulation for an assumed 10 meter aperture. The atmospheric coherence length is a parameter of the simulation, which we use to adjust the average Strehl ratio of the PSF delivered by the AO correction. Our particular implementation of Kolmogorov turbulence employs the Fourier-based autoregressive algorithm of \cite{Srinath:15}; an example aberrated wavefront in the pupil plane generated with this method is shown in Fig.~\ref{fig:wfs}a. However, Fourier-based turbulence models typically under-represent low-order WFEs \cite{Johansson:94}. Additionally, the Kolmogorov power spectrum does not account for other sources of low-order error such as non-common path aberrations (NCPAs) and mechanical vibrations. As such, extra low-order phase errors can be added to make WFE more realistic. In this work, we allow for the injection of extra tip-tilt error by applying a Gaussian-distributed random displacement to the PSF in the waveguide input focal plane. The magnitude of this displacement is a parameter of the simulation that can be adjusted. Static WFEs, such as those caused by misalignment or imperfect optical prescriptions (giving rise to, say, coma or astigmatism), are neglected.  

\subsection{Adaptive optics}\label{sec:ao} 
The turbulence-aberrated wavefronts as simulated above are partially corrected using an AO model built with HCIPy \cite{hcipy}, a high-contrast imaging package written for Python. The general trends in our results will be independent of pupil geometry, but for the purpose of this work we assume the Keck pupil. We assume a deformable mirror (DM) with $30 \times 30$ actuators across the pupil, in line with current ``extreme AO'' systems and similar to the $\sim$1,000-element DM which will be introduced to Keck as part of the phase II of the KPIC project~\cite{Jovanovic:20}. The control law is handled using a basic leaky-integrator scheme. Propagation from the pupil to focal plane is accomplished using HCIPy's Fraunhofer propagator.  Example PSFs from our model in the presence and absence of turbulence-induced WFE are shown in Figs.~\ref{fig:wfs}b and \ref{fig:wfs}c. Finally, for each tested lantern entrance diameter, we adjust the focal ratio of the optical system in order to maximize the spatial overlap between the PSF and the lantern modes, ensuring optimal coupling efficiency.

\subsection{Beam-shaping (PIAA) optics}\label{sec:piaa_method}
PIAA optics remap the amplitude distribution of light in the pupil plane, in order to promote a more amenable amplitude distribution (e.g. for fiber injection or coronagraphy) in the focal plane. This remapping is usually accomplished by a pair of lenses (or mirrors) placed after the wavefront corrector. In the context of SMF injection, the target amplitude distribution is usually selected to follow a truncated, circularly symmetric Gaussian profile. This profile removes the Airy ring, created by diffraction from the pupil mask, and promotes a Gaussian-like PSF that matches well with the Gaussian-like fundamental mode of SMFs (\cite{Jovanovic:17}; also see the first panel of Fig.~\ref{fig:wfs}g for an example of the fundamental mode). The action of such beam-shaping optics is shown in the second row of Fig.~\ref{fig:wfs}: Fig.~\ref{fig:wfs}d shows the Keck pupil post-apodization, while Figs. \ref{fig:wfs}e and \ref{fig:wfs}f show apodized PSFs with and without turbulence-induced aberrations. Without beam-shaping optics, coupling efficiencies into SMFs for optical systems with Keck-like pupils (secondary obstruction ratio of $\sim$30\%) would be limited to $\sim$65\%; with such optics, coupling efficiencies can reach upwards of 80\% \cite{Jovanovic:17, Calvin:21}. In this work we consider how these optics may also be applied to lantern-based fiber injectors. We computed the required lens profiles corresponding to the desired Gaussian amplitude distribution by numerically solving a differential equation for the lens surfaces, obtained in the geometric optics limit \cite{Guyon:03}. When we use beam-shaping optics in our simulations, we propagate wavefronts through lenses with the derived lens profiles, after correction from the DM but before propagation to the focal plane. For free-space propagation between the two lenses, we use HCIPy's angular spectrum propagator.

\subsection{Photonic lantern}\label{sec:lant_method}
After propagation to the focal plane, light is injected into a PL. In this work, we only consider simple lanterns of the type shown in Fig.~\ref{fig:lant}, where the transverse structure of the transition zone at any point along its length is just a linearly scaled-down version of the lantern's output (i.e. isolated single-mode cores in a common cladding). The scale difference between the output and input ends we term the ``taper factor''. Such lanterns deviate slightly from real lanterns in that the claddings of real lanterns do not necessarily maintain circularity as they taper down; however, they are a good first approximation. Due to the construction method, lanterns formed by tapering fibers or fiber bundles may also have a ``lead-in'' length of MMF-like waveguide at their input end, which we simulate in our model by extending the tapered end of the transition zone. 
\\\\
The most straightforward way to simulate lantern throughput is to numerically propagate injected light through the lantern and compute the power in the output lantern modes. However, numerical propagation is computationally expensive. This expense is multiplied in the presence of stochastic processes such as turbulence, where numerical propagation must be repeated for multiple realizations of the input electric field in order to obtain a statistical average, and additionally compounded during parameter space explorations (e.g. over WFE or wavelength, as done in this work). To simplify and accelerate our simulations, we split throughput into two parts: coupling efficiency, which reflects the initial loss of light during coupling into the lantern entrance, and transition efficiency, which reflects losses that occur during propagation through the lantern. Previous works have shown that transition efficiencies can be as high as $\sim$97\% \cite{Trinh:13,Leon-Saval:14}; a necessary but insufficient condition for this to occur is to have at least as many single-mode outputs (or ``ports") at the lantern exit as guided modes at the lantern entrance \cite{Birks:15}. Assuming negligible transition effects, the input coupling efficiency of the lantern can be taken as a proxy for overall lantern throughput, bypassing the need to simulate the lantern transition and saving on computation time. In this work, lantern coupling efficiencies are generally computed under the assumption that the number of ports equals the number of modes and transition losses are negligible.
\\\\
Coupling efficiencies on their own are quick to simulate. We only need to compute a basis of “lantern modes’’: a set of orthogonal, linearly independent field distributions at the lantern entrance which span the space of bound electric field distributions when propagated to the lantern output. Then, the coupling efficiency into a lantern entrance for a given input electric field is simply the total amount of power coupled into all of the lantern modes, which we can compute via overlap integrals between the input field and the lantern modes. The overlap integral $\eta$ for two scalar complex fields $E_1$ and $E_2$ is given by 
\begin{equation}
    \eta = \dfrac{ \left| \int E_1^* E_2 dA \right|^2 }{\int \left| E_1 \right|^2 dA \int \left| E_2 \right|^2 dA }.
\end{equation}
For large lantern taper factors (small residual SMF cores at the lantern's entrance),  the lantern modes approach the LP$_{lm}$ modes: analytically known eigenmodes of propagating waves in a circularly symmetric step-index fiber. Example amplitude distributions for the first 19 LP modes are shown in Fig.~\ref{fig:wfs}g (rotated variants of modes $l \neq 0$ are not shown).
\\\\
Later in this work we compute lantern throughputs (rather than input coupling efficiencies), and extend lantern characterization over regimes where lantern operation has been considered ``inefficient'' and the above approach breaks down. In these regimes, we need to apply full numerical beam propagation to connect the electric field distributions at the lantern input and output. In order to simulate the propagation of light through PLs, we have built ``Lightbeam'' \cite{mybpm}: a numerical package which uses the ``finite-differences beam propagation method'' (FD-BPM) algorithm to numerically and implicitly solve the paraxial Helmholtz equation in the weakly guiding regime. Our particular variant of FD-BPM is based on the algorithm presented in \cite{bpm}, which is accurate to 4$^{\rm th}$ order transversely, to 2$^{\rm nd}$ order longitudinally, and works on an adaptive mesh. We have tested this code against RSoft BeamProp, a commercial package for optical propagation, with consistent results. We have also verified that our code reproduces the expected behavior for single and few-mode circularly symmetric step-index fibers. 

\subsection{Tested configurations}
All simulated lanterns are assumed to have a fluorine-based glass jacket and a fused 
silica cladding with an index contrast of $5.5\times10^{-3}$, materials typically used to construct lanterns (private communication with S. Leon-Saval). We consider lantern operation across the astronomical $y$ and $J$ bands (0.97--1.35 $\upmu$m), where the Strehl ratio of typical AO systems is lower and there is a stronger need for PLs. The embedded SMF cores are assumed to have similar specifications to OFS ClearLite 980 16 fiber, which has a core-cladding index contrast of $8.8\times10^{-3}$ and a nominal core diameter of 4.4 ${\upmu}$m. This fiber was chosen due to its single-mode cutoff at 970 nm, consistent with our chosen wavelength range. We additionally assume linear lantern taper profiles throughout this work; other geometric parameters are adjustable.
\\\\
We use this model to assess lantern performance as a function of entrance diameter, wavelength, and WFE, as well as the length of the lantern's lead-in waveguide and taper factor. We also compare the performance of SMFs and 3-, 6-, and 19-port lanterns with and without beam-shaping optics. We have chosen to consider 3- and 6- port lanterns because their low port counts make them more applicable for use in high-resolution spectrometers, which require many pixels and lack room for many fibers. We additionally consider the 19-port lantern because such devices are already in use (e.g. GNOSIS~\cite{Trinh:13}, PRAXIS~\cite{ellis2018a}). Additionally, these port numbers were selected because they reflect how the number of guided modes increases as core diameter is scaled up and/or wavelength is scaled down.  

\section{Results}\label{sec:res}
In this section, we characterize lantern performance as a function of Strehl ratio, operating wavelength, and geometric factors including lantern entrance diameter, taper factor, and lead-in waveguide length. Sections~\ref{sec:coupvsstrehl} through \ref{sec:TT} characterize lantern performances in terms of coupling efficiency into the lantern entrance, which serves under most circumstances as a good proxy for lantern throughput. In Section~\ref{sec:fullprop}, we use full numerical beam propagation to compute proper throughputs and extend the results of prior sections.

\subsection{Coupling efficiency vs. Strehl ratio}\label{sec:coupvsstrehl}
\begin{figure}[h!]
\centering\includegraphics[width=\textwidth]{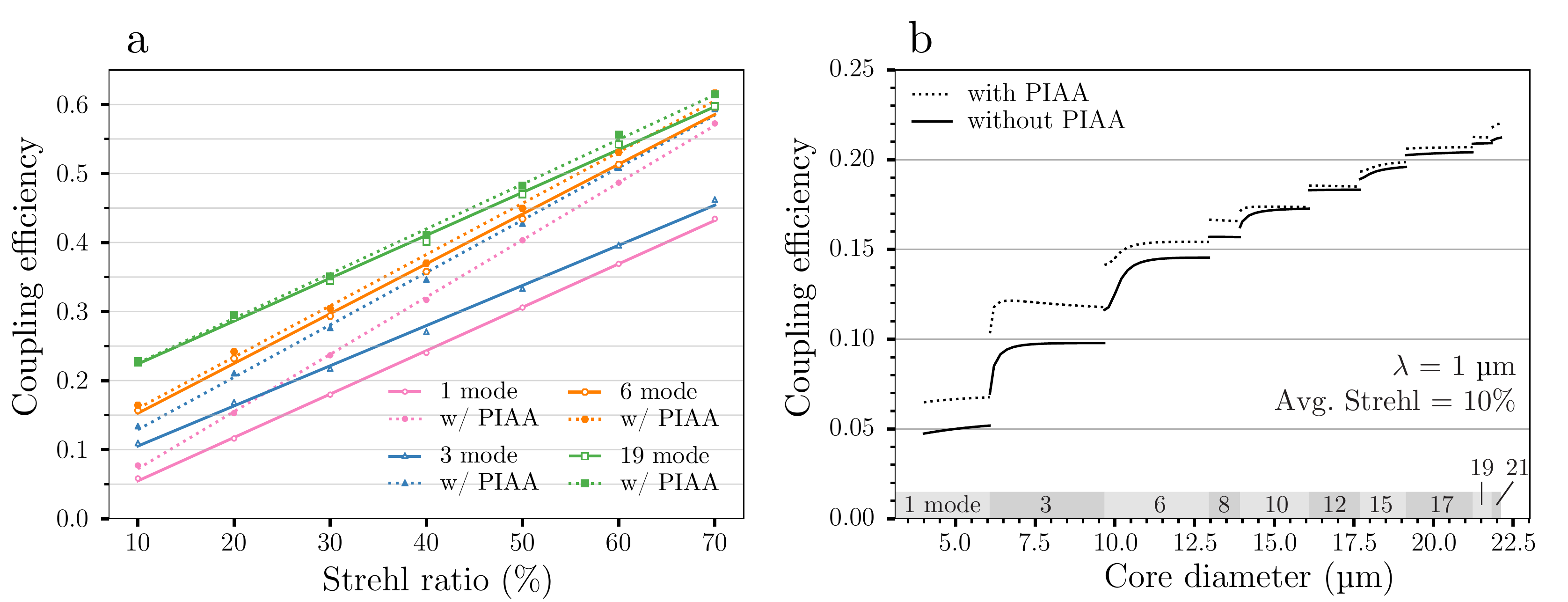}
\caption{a: coupling efficiency comparison for 1-, 3-, 6-, and 19-mode lanterns against Strehl ratio at $\lambda =$ 1~$\upmu$m, with and without beam-shaping (PIAA) optics. Lantern entrances were approximated as step-index fibers. The specified mode counts were selected because they reflect how the number of LP modes increases with core diameter. PSF degradation was simulated using standard Kolmogorov turbulence. b: Mean fiber coupling efficiency in the presence and absence of beam-shaping optics, at a wavelength of 1~$\upmu$m for an average Strehl ratio of 10\%. The number of available LP modes is annotated at the bottom. } \label{fig:vs_strehl_vs_cd}
\end{figure}

With the index contrast of the lantern fixed, the primary factor that governs lantern coupling efficiency is the diameter of the lantern's input waveguide, which sets the number of supported modes at the lantern's entrance and thus determines the lantern's ability to accept aberrated light. We present coupling efficiencies at $\lambda=$ 1~$\upmu$m into the input ends of lanterns for various entrance diameters (and hence mode counts) as a function of Strehl ratio in Fig.~\ref{fig:vs_strehl_vs_cd}a. The Strehl ratio is computed from the intensity average of 100 turbulence-aberrated PSF realizations. We additionally adjust focal ratio to maximize the overall coupling efficiency at each core diameter (re-optimizing the focal ratio when beam-shaping is included in the optical system). We assume that the embedded SMF cores are small after tapering (large taper factor) and that the lantern entrance is circular, so that the lantern modes can be taken as the LP modes. Other geometric parameters such as lead-in length are for now irrelevant; we consider them again in Section \ref{sec:fullprop}. WFE was solely generated from Fourier-based Kolmogorov turbulence (no extra tip-tilt was added; we consider extra tip-tilt later in \S\ref{sec:TT}).  Figure~\ref{fig:vs_strehl_vs_cd}a indicates that coupling efficiencies increase linearly with Strehl ratio independent of mode count, with the largest mode count lantern consistently outperforming all other configurations. Performance gains from PLs over SMFs are also evident across the entire range of tested Strehl ratios: even at 10\% Strehl, the simplest three-mode lantern outperforms direct injection into an SMF by a factor of 2, netting a throughput of $\sim$10\%. In the absence of beam-shaping optics, six-mode lanterns offer even greater gains, reaching a $\sim$15\% increase over both SMFs and three-port lanterns at 70\% Strehl ratio.

\subsection{Effect of lantern size}\label{sec:lantsize}
In Fig.~\ref{fig:vs_strehl_vs_cd}b, we extend the previous calculation over a range of core diameters, this time fixing turbulence to give an average 10\% Strehl at the operating wavelength of 1~$\upmu$m. As its diameter increases, the number of guided modes at the lantern's entrance increases in a stepwise fashion, resulting in discontinuous jumps in the lantern coupling efficiency. Importantly, the number of guided modes (annotated at the bottom of the plot) does not increase one at a time: modes usually appear in groups of 2 or 3. Figure~\ref{fig:vs_strehl_vs_cd}b shows that the most significant gains in coupling efficiency occur in the 1- to 3-mode transition and the 3- to 6-mode transition at 10\% Strehl, beyond which increases in lantern size and modality yield diminishing returns. However, if the increases in instrument complexity and exposure times are not a concern, it may still be beneficial to adopt a high-mode count lanterns: as seen in Fig.~\ref{fig:vs_strehl_vs_cd}, the coupling efficiency of a 19-port lantern is around double that of a 3-port lantern at 10\% Strehl.
\begin{figure}[h!]
    \centering
    \includegraphics[width=\textwidth]{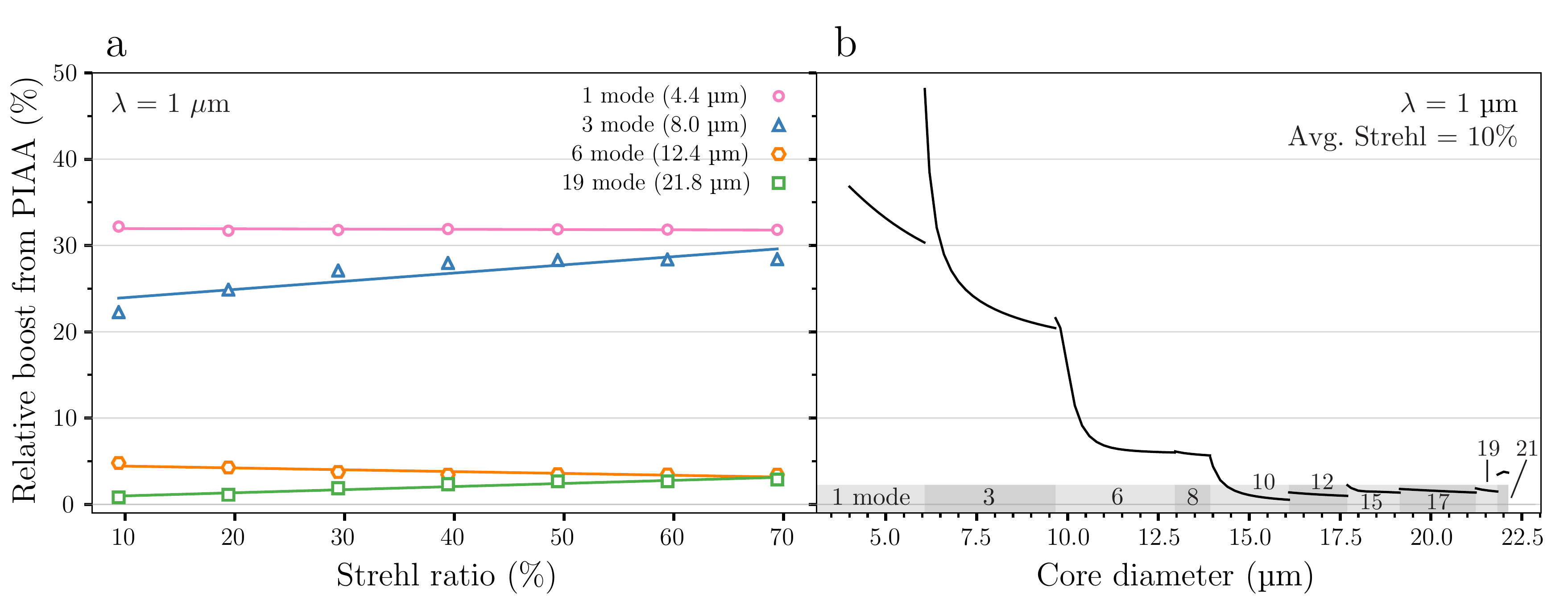}
    \caption{a: relative coupling efficiency boost offered by Gaussian beam shaping optics as a function of Strehl ratio, for various lantern entrance diameters. Large lanterns show less gain over all tested Strehl ratios. b: relative coupling efficiency boost from beam shaping optics as a function of lantern entrance diameter. We approximate the lantern entrance as a step-index MMF. As diameter increases, the boost offered by beam shaping optics decreases, dropping below 10\% in the 6-mode region.}
    \label{fig:boost}
\end{figure}

\subsection{Effect of beam-shaping (PIAA) optics}\label{sec:piaa}
Figure~\ref{fig:vs_strehl_vs_cd} shows the absolute coupling gains from beam-shaping optics as a function of Strehl ratio and lantern entrance diameter. We recast these results in terms of the relative gains offered by beam-shaping optics in Fig.~\ref{fig:boost}. As indicated in panel a, we see that beam shaping optics offer a significant $\sim$30\% boost to SMF coupling, consistent with results from \cite{Jovanovic:17,Calvin:21}. We additionally see that these beam-shaping optics also offer a significant $\sim$25\% boost to coupling into a 3-port lantern, but offer negligible $\sim$5\% gain for a 6-port lantern.  Consequently, we find that the 3-port PL with beam-shaping performs almost as well as a 6-port lantern over the entire range of Strehl ratios (see Fig.~\ref{fig:vs_strehl_vs_cd}a). Figure~\ref{fig:boost}b makes a similar argument, showing that as lantern entrance diameter (and the number of supported modes) increases, the relative boost afforded by beam-shaping optics drops, declining below 5\% beyond the 6-mode region, with negligible gains for lanterns with 10+ ports. The reason for this will be described in Section~\ref{sec:disc}.
\\\\
The bump in relative boost in the 21-mode region shown in Fig.~\ref{fig:boost}b can be traced to the LP$_{13}$ mode, which in the 21-mode region is very close to its cutoff wavelength and thus extends further into the cladding. This mode shape matches well with the amplitude distribution produced by our beam-shaping optics; a similar effect can be seen at the beginning of the 3 mode region in the same figure, where the relative boost initially spikes because the LP$_{11}$ mode is close to its cutoff wavelength and thus has a larger spatial extent. Such modes can be more sensitive to losses (e.g. due to bending \cite{fiber}) so it is unclear if these coupling gains are realizable or if they are an artifact of our idealized simulations.  

\subsection{Chromatic behavior}\label{sec:chrom}
\begin{figure}[h!]
\centering\includegraphics[width=10cm]{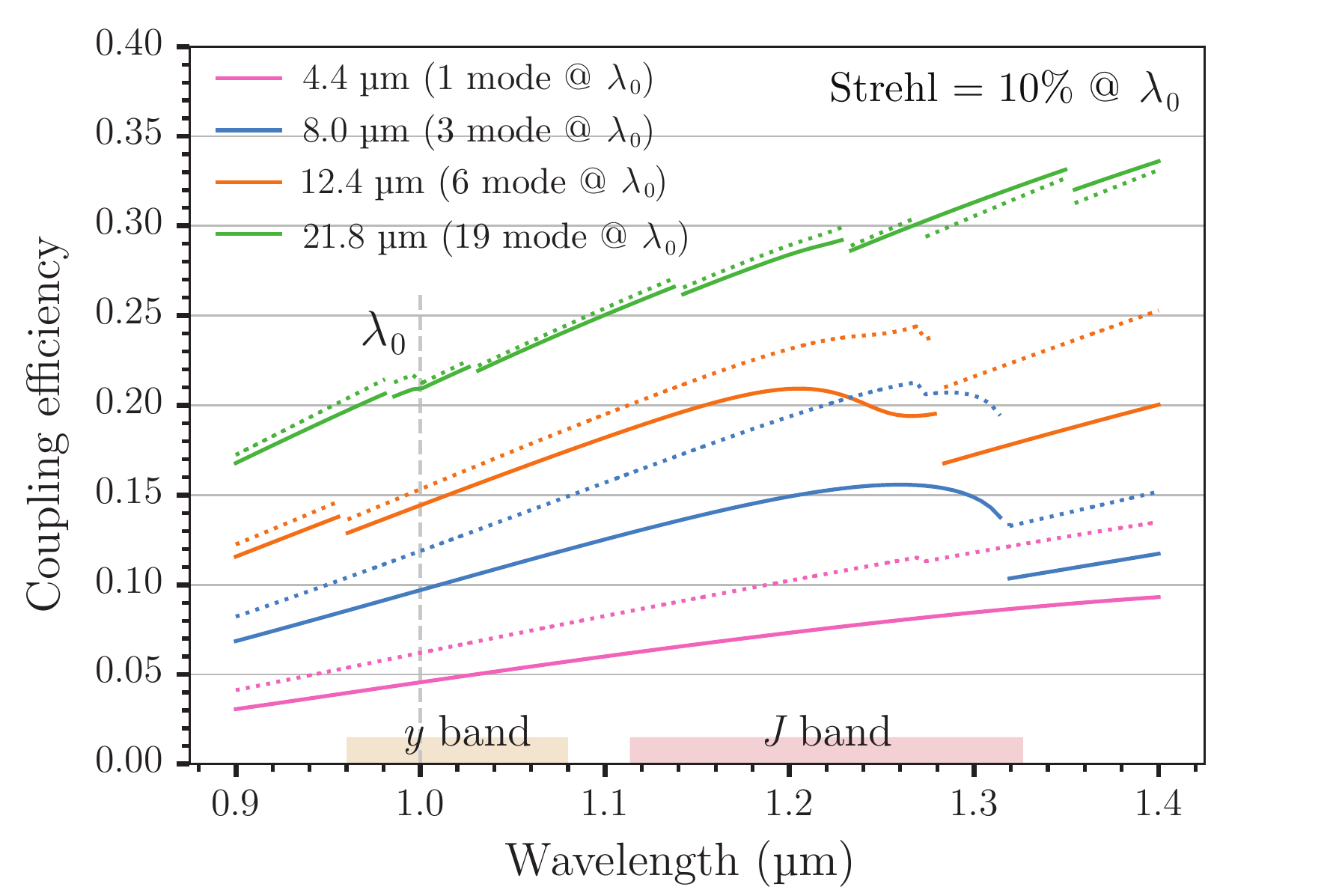}
\caption{Throughput comparisons for the lantern configurations in Fig.~\ref{fig:vs_strehl_vs_cd}a, now as a function of wavelength. Simulated atmospheric conditions were fixed for an average of 10\% Strehl at $\lambda$ = 1~$\upmu$m. Dotted lines show the coupling efficiencies with beam-shaping (PIAA) optics. For each curve, the focal ratio is optimized at the nominal core diameter assuming a wavelength of 1~$\upmu$m.} \label{fig:chrom}
\end{figure}

We extend the results from Fig.~\ref{fig:vs_strehl_vs_cd}a over wavelength in Fig.~\ref{fig:chrom}, again fixing atmospheric turbulence to give PSFs with an average 10\% Strehl at the nominal operating wavelength of 1~$\upmu$m. Here, we fix the number of PL ports to be equal to the number of modes at the PL entrance, at the reference wavelength $\lambda = 1$ $\upmu$m. With the exception of the 19-port lantern, this assumption ensures that the lanterns will have at least as many output ports as entrance modes over the $y$ and J bands. Unsurprisingly, we see from Fig.~\ref{fig:chrom} that configurations with larger lanterns consistently outperform those with smaller lanterns across the entire range of tested wavelengths. Additionally, as wavelength increases, phase error decreases, and coupling efficiency into lanterns typically increases for all entrance diameters, though there are dips and discontinuous drops as the guided modes at the lantern entrance change shape and eventually disappear. 
The discontinuous drops in coupling efficiency seen in Fig.~\ref{fig:chrom} indicate transition wavelengths across which the number of guided modes at the lantern entrance changes. Importantly, near such wavelengths, we no longer expect coupling efficiency to be an accurate proxy for lantern throughput. The reasoning is as follows: as wavelength increases past a transition wavelength, lantern modes at the waveguide entrance that were originally guiding now become weakly radiative. Normally, we expect that such modes would not be able to carry power; however, if the characteristic length scale for radiative attenuation is small compared to the length of the lantern lead-in, losses are negligible and these radiative modes can still transfer power into the lantern cores. This effect is relatively unimportant for large lanterns (e.g. a few percent difference in coupling efficiency for the 19-port lantern in Fig.~\ref{fig:chrom}) since such lanterns will distribute power over many modes. However the effect can be significant for few-mode lanterns. In Section~\ref{sec:fullprop}, we apply numerical beam propagation to compute throughputs of few-port lanterns in these transition regimes.

\subsection{Tip-tilt wavefront error}\label{sec:TT}
\begin{figure}[h!]
\centering\includegraphics[width=\textwidth]{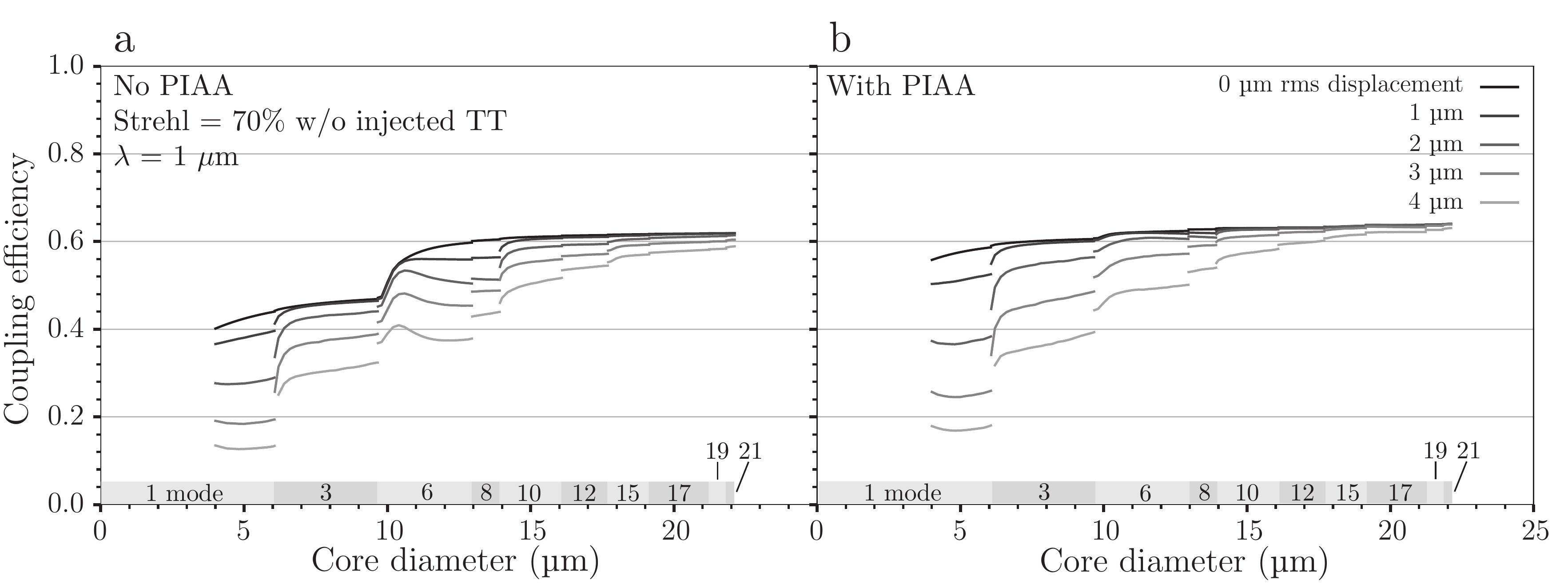}
\caption{Similar to Fig.~\ref{fig:vs_strehl_vs_cd}b, but the atmospheric coherence length is fixed to give an average Strehl of 70\%, after which tip-tilt WFE in the form of random displacements in the focal plane is injected to wavefront realizations. This tip-tilt is added with the intention to approximate instrumental tip-tilt and compensate for potential underestimation of low-order error in our turbulence model.} \label{fig: TT1}
\end{figure}

\begin{figure}
\centering\includegraphics[width=10cm]{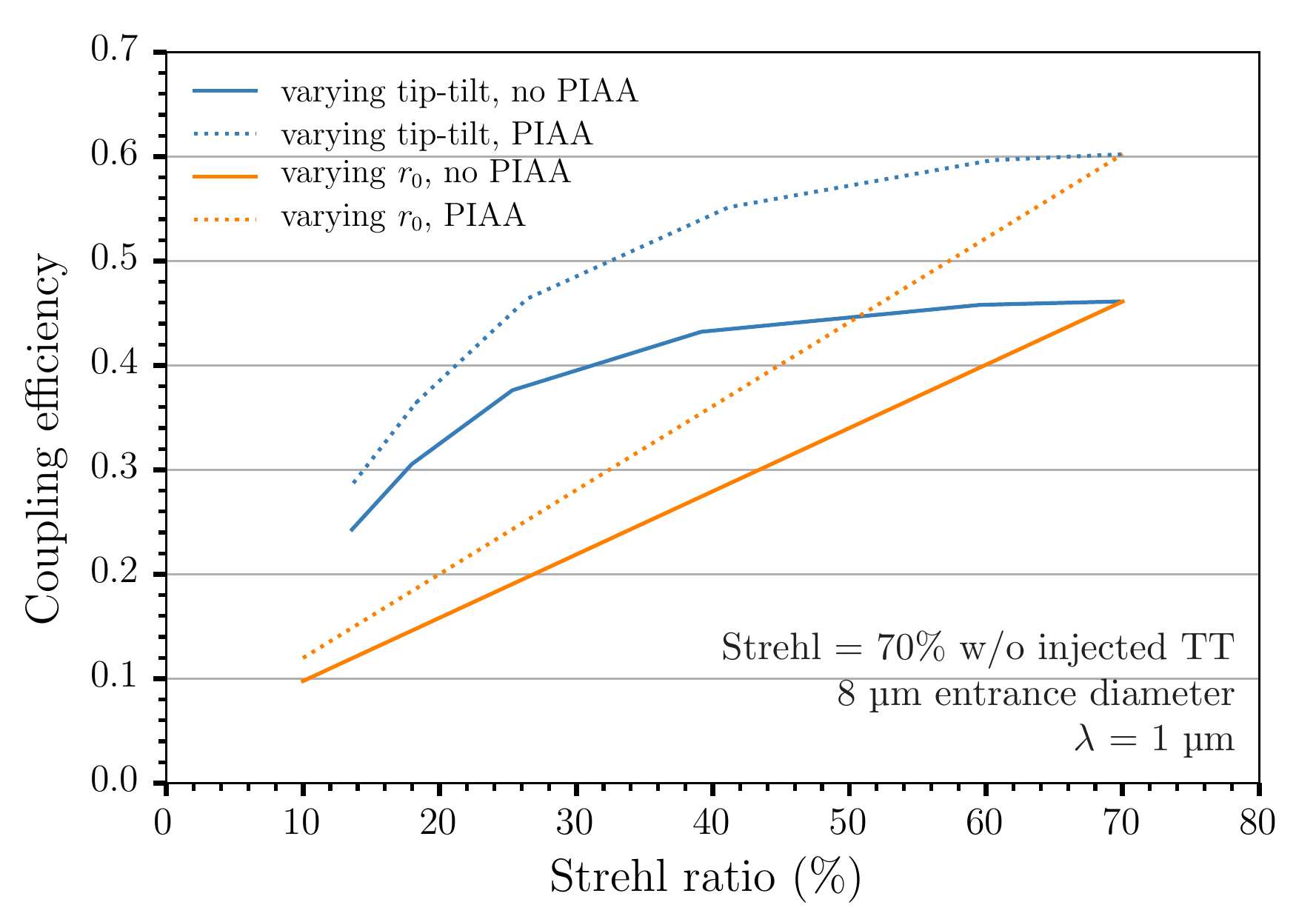}
\caption{Coupling efficiency vs Strehl ratio into an 8~$\upmu$m diameter (3 mode) fiber for aberrated PSFs. Aberrations can be introduced in multiple ways; we consider two. In the first we assume only single-layer atmospheric turbulence and vary the coherence length $r_0$. In the second we degrade the PSFs by fixing some $r_0$ and then injecting additional tip-tilt error in the form of random displacements in the focal plane. In this case, we fixed the $r_0$ to give 70\% average Strehl before injecting tip-tilt. Both with and without beam-shaping (PIAA) optics, MMFs are better able to capture light degraded via the second method.    } \label{fig:TT2}
\end{figure}

To make our WFE more realistic, we can inject extra tip-tilt in addition to the existing WFE from Kolmogorov turbulence. Similar to Fig.~\ref{fig:vs_strehl_vs_cd}b, Fig.~\ref{fig: TT1} plots the mean coupling efficiency for a PL as a function of entrance diameter, though now under varying amounts of injected tip-tilt error. The focal ratio of the optical system was optimized to match the size of the lantern modes at each core diameter prior to tip-tilt injection. Scenarios without beam-shaping optics (panel a) and with such optics (panel b) are considered. The atmospheric coherence length is tuned to give an average 70\% Strehl ratio at a wavelength of 1~$\upmu$m (actual Strehl of the intensity-averaged image, after tip-tilt injection, will be lower). Figure~\ref{fig: TT1} shows that coupling degrades as both the amount of injected tip-tilt increases and as entrance diameter decreases, suggesting that the reduction in coupling performance due to the injected tip-tilt ultimately depends on the relative amount of tip-tilt displacement compared to the size of the lantern entrance. Coupling appears mostly unaffected so long as the RMS tip-tilt displacement is smaller than the lantern entrance radius by more than a factor of 2. 
\\\\
We present an alternative perspective on lantern performance in the presence of tip-tilt in Fig.~\ref{fig:TT2}, comparing coupling efficiency against Strehl ratio for a lantern entrance diameter of 8 $\upmu$m in two distinct cases: either tuning Strehl by adjusting atmospheric coherence length or by adjusting the RMS displacement of injected tip-tilt error. Figure~\ref{fig:TT2} indicates that coupling efficiencies are comparatively better in the latter case, implying that few-mode fibers and PLs are resilient against tip and tilt: for a given amount of WFE, coupling efficiency of the beam into the fiber or lantern improves when there is more power in tip-tilt aberrations vs pure Kolmogorov turbulence. This holds whether or not beam-shaping optics are used. With this in mind, we see that the results of \S\ref{sec:coupvsstrehl}-\S\ref{sec:chrom} and later \S\ref{sec:fullprop}, which do not include additional tip-tilt, are lower bounds on the coupling gains that can be realized by lanterns in real optical systems. The degree of this effect will depend on lantern size and the specific characteristics of instrumental tip-tilt.

\subsection{Full beam propagation}\label{sec:fullprop}
\begin{figure}[h!]
    \centering
    \includegraphics[width=10cm]{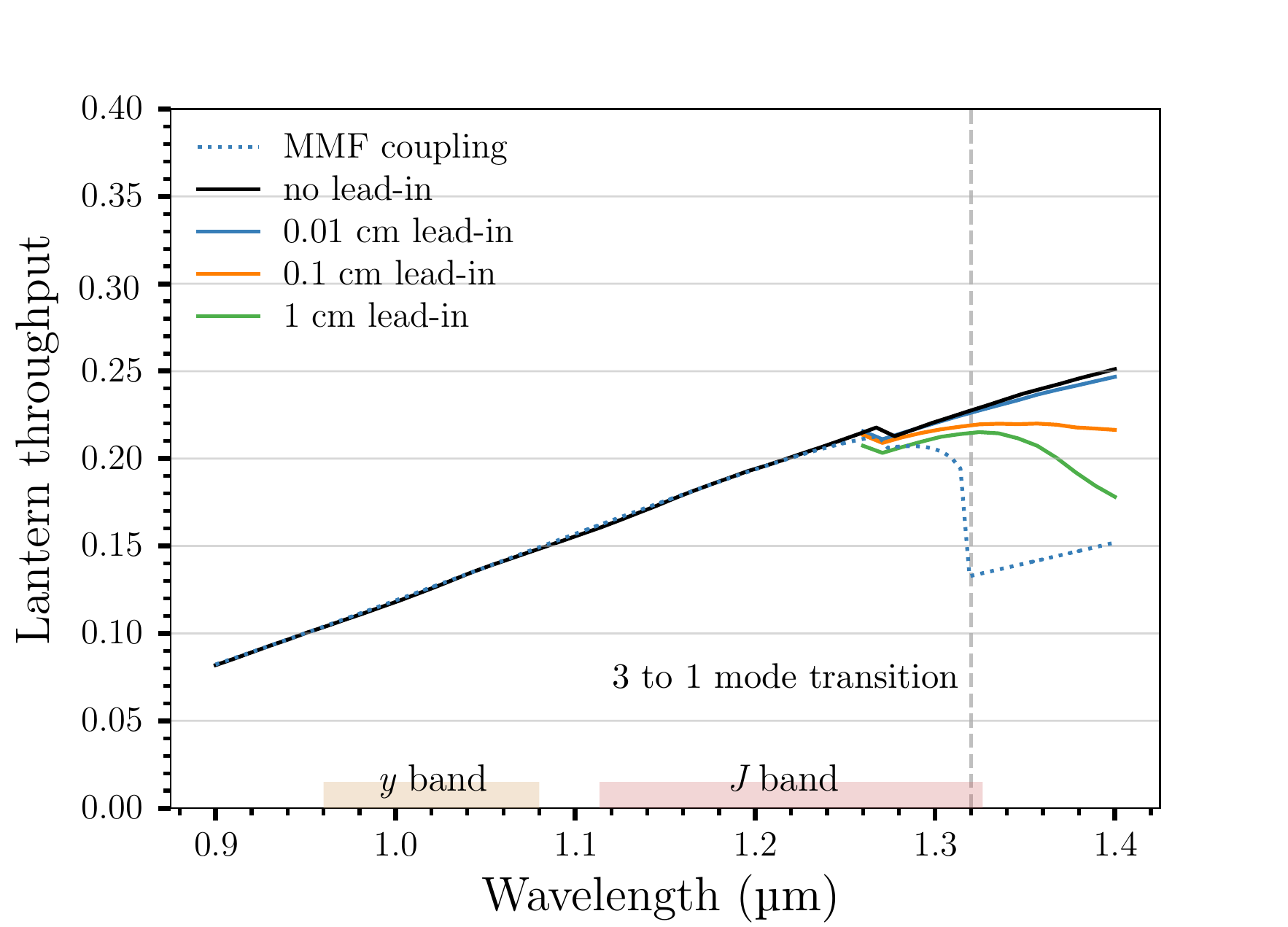}
    \caption{Throughput of an 8 $\upmu$m diameter lantern against wavelength, for various lantern lead-in lengths. WFE here is wholly generated from Kolmogorov turbulence with an average 10\% Strehl. The dotted curve denotes coupling efficiency into an 8 $\upmu$m diameter circular step-index fiber; all other curves denote lantern coupling efficiencies and were generated using beam propagation. Beyond 1.32 $\upmu$m, the step-index fiber becomes single-mode, and coupling efficiency drops precipitously. However lanterns with short lead-in lengths do not experience such a drop due to radiative modes. Note that throughputs for non-zero lead-in lengths were only computed and plotted for $\lambda>1.26$ $\upmu$m to save on computation time.}
    \label{fig:leadin}
\end{figure}

\begin{figure}[h!]
    \centering
    \includegraphics[width=10cm]{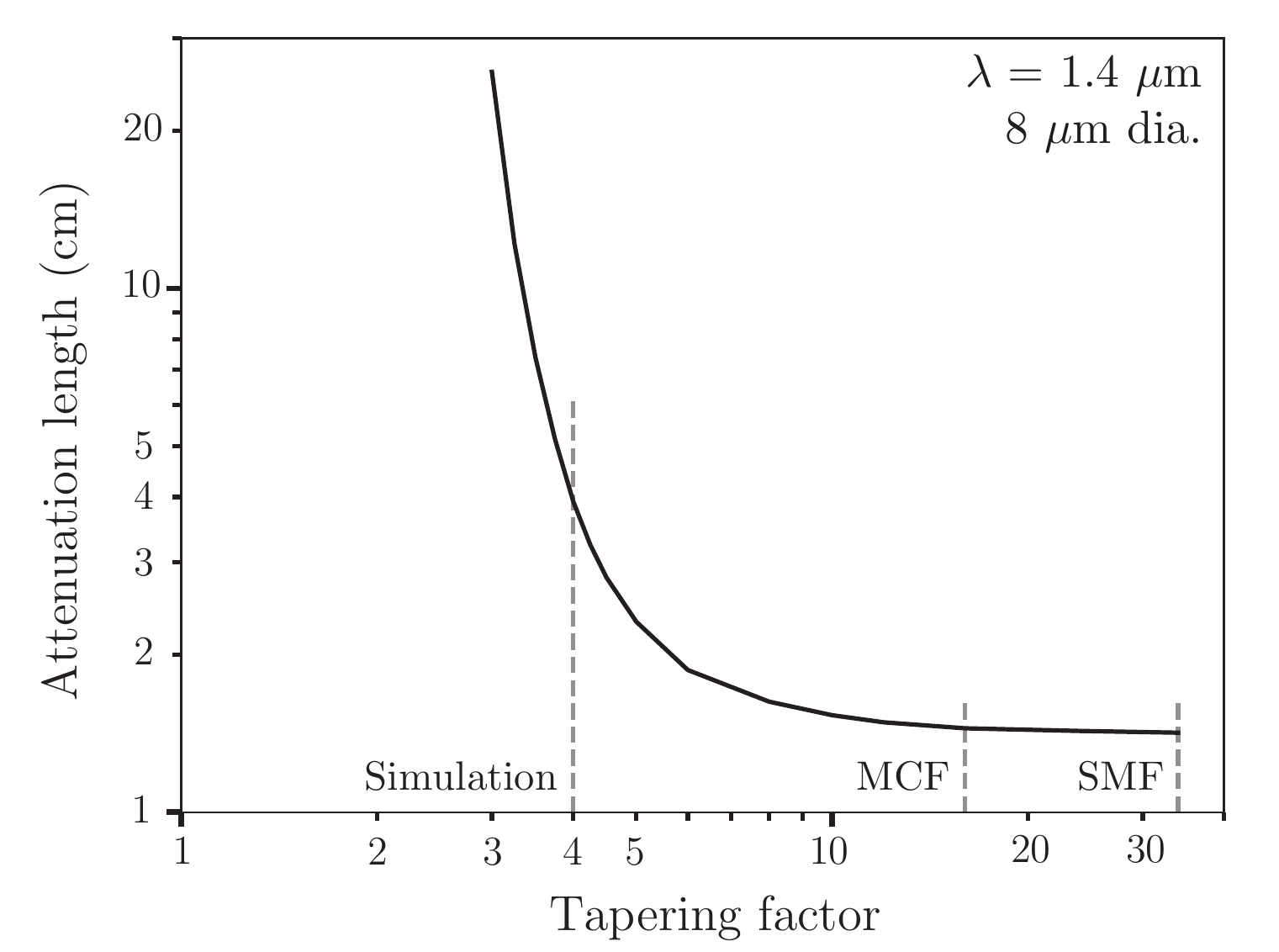}
    \caption{Attenuation strength vs. taper factor for one of the radiative LP$_{11}$-like modes ($\lambda$ = 1.4~$\upmu$m) in a lead-in waveguide for a 3-port, 8~$\upmu$m diameter lantern, computed with FEMSIM \cite{femsim}.  Attenuation strength is quantified as the length over which the total power of the radiative LP$_{11}$-like mode drops by a factor of $e$.  The required taper factors to construct the assumed lantern from a bundle of SMFs (125 $\upmu$m diameter cladding) and a MCF (125 $\upmu$m diameter cladding overall), as well as the taper factor of the lantern simulated in Fig.~\ref{fig:leadin}, are annotated with vertical dashed lines.}
    \label{fig:atten}
\end{figure}

Because beam propagation is more computationally expensive than the prior coupling calculations, we apply this method only to study the simplest 3-port lantern; the general trends from this section will be extensible to larger lanterns.
Figure~\ref{fig:leadin} presents the throughput (not input coupling efficiency) for a 3-port lantern with an 8~$\upmu$m entrance diameter with beam-shaping optics (black curve). In throughput calculations, geometric factors such as lead-in length and taper factor now become relevant; we assume a $4 \times $ taper factor, and no lead-in waveguide. To generate the throughput curve in Fig.~\ref{fig:leadin}, we have numerically computed the lantern modes at the waveguide entrance by backpropagating the fundamental modes of the SMF cores from the lantern output and orthogonalizing the resulting basis. These modes are LP-like, but will no longer necessarily be guiding within the lantern entrance if there are more lantern cores than propagating modes at the lantern's MMF-like input. Calculation of the lantern modes allows us to retain our modal analysis approach and compute lantern throughput from the input electric field distribution at the lantern entrance. As a result, we bypass the need to forward-propagate multiple realizations of the input field through the guide, saving on computation time. As seen in Fig.~\ref{fig:leadin}, lantern throughput agrees well with the coupling efficiency into a similarly sized step-index fiber until the propagating wavelength approaches the $3\rightarrow 1$ mode transition wavelength at $\sim 1.32$ $\upmu$m. Beyond 1.32 $\upmu$m, the lantern throughput exceeds step-index fiber coupling efficiency by almost a factor of 2. This is because an 8 $\upmu$m diameter step-index fiber with our assumed core and cladding materials becomes single-mode beyond $\lambda = 1.3$ $\upmu$m; however, an 8 $\upmu$m diameter lantern entrance retains weakly attenuating radiative modes which can help transfer power to the lantern cores.
\\\\
As such, over transition wavelengths where radiative lantern modes are active, the length of the lead-in waveguide becomes a relevant factor in lantern design. To show this effect, Fig.~\ref{fig:leadin} plots throughput curves for lanterns that are similar to our 3-port configuration with the exception of differing lead-in lengths (note these throughputs were only computed for $\lambda > 1.26$ $\upmu$m to save on computation time). Throughputs are now computed using the ``brute-force'' method of numerically propagating all input field realizations to the lantern's output. Note that we can no longer use backpropagation when attenuation through the lantern is non-negligible (as can be the case for lead-in waveguides): the departure of light from the simulation zone makes the operation of numerical propagation non-invertible. Figure~\ref{fig:leadin} verifies that longer lead-in lengths result in greater dampening in lantern throughput at wavelengths beyond the transition wavelength of $1.32$ $\upmu$m, due to increased attenuation of the radiative modes (note that this attenuation is being captured by the numerical beam propagation itself, not some assumed fiber model). The magnitude of attenuation also increases as the propagation wavelength increases beyond the transition wavelength. 
\\\\
The reduction in throughput due to attenuation of the radiative modes will also depend on the lantern taper factor, which sets the degree by which the embedded SMF cores in the lantern shrink from output to input. Figure~\ref{fig:atten} plots the degree of attenuation for the LP$_{11}$-like radiative mode in the lead-in waveguide of a 3-mode (8~$\upmu$m entrance diameter) lantern, as a function of taper factor $t$; since the SMF cores at the output end of the lantern are 4.4~$\upmu$m in diameter, the diameter of the tapered-down cores embedded in this lead-in waveguide can be computed as 4.4~$\upmu$m / $t$. The propagation wavelength is taken to be 1.4 $\upmu$m. Attenuation is quantified in terms of the attenuation length $d$, the distance over which power in the radiative mode drops by a factor of $e$. The attenuation length $d$ can be computed from
\begin{equation}
    d = \dfrac{\lambda}{4 \pi {\rm Im} \left(n_{\rm eff}\right)}
\end{equation}
where $n_{\rm eff}$ is the (complex) effective index of refraction for the radiative mode and ${\rm Im}$ denotes taking the imaginary part. In turn, we compute the effective index using FemSIM~\cite{femsim}, a finite-element mode solver that can compute the effective indices of radiative modes. From Fig.~\ref{fig:atten}, we see that the attenuation of this radiative mode increases as taper factor decreases and the embedded SMF cores at the lantern entrance become larger. The dependence between the attenuation length and taper factor is asymptotic, climbing rapidly to infinity as the embedded SMF cores at the lantern's entrance approach the minimum required size to become guiding. Finally, we note from Fig.~\ref{fig:atten} that the attenuation length for a 3-port lantern with a taper factor of 4 is $\sim$3 cm. This is order-of-magnitude consistent with Fig.~\ref{fig:leadin}, computed using our numerical beam propagator, where we see that the 3-port, 4$\times$ tapered lantern begins to show significant attenuation at $\lambda = 1.4$ $\upmu$m when the lead-in length reaches 1 cm.

\section{Discussion/Future directions}\label{sec:disc}
Our results from Fig.~\ref{fig:vs_strehl_vs_cd}a recover the linearly increasing trend of coupling efficiency with Strehl ratio, consistent with prior works~\cite{Jovanovic:17}. This is sensible: the Strehl ratio is directly proportional to peak PSF intensity, which in turn is directly proportional to the amount of PSF power that couples into the lantern modes. The coupling boost provided by the beam-shaping optics also increases with Strehl ratio, since beam-shaping optics only correct for average pupil illumination and not WFE. The boost is greatest for SMFs, which in conjunction with beam-shaping actually approach the performance of 6-port lanterns at high Strehl. This behavior is expected: at high Strehl, beam-shaping moves the bulk of light at the focal plane into the PSF core, promoting coupling into the single Gaussian-like mode of SMFs. In contrast, larger lanterns have access to other higher-order modes which can couple light away from the PSF core, such as light in the Airy rings, reducing the utility of beam-shaping. In the case of the 6-port lantern, the presence of the LP$_{02}$ mode allows such lanterns to efficiently couple light from the Airy ring. However, in the low-Strehl regime PLs are still preferable. This result is reaffirmed in the presence of increased tip-tilt error: the extra size and modality of lanterns makes their coupling efficiencies less sensitive to displacement of the PSF core. In fact, Fig.~\ref{fig:TT2} shows that for tip-tilt-varied WFE, lantern coupling efficiency scales roughly with the square root of the Strehl ratio. In contrast, coupling scales linearly with Strehl when WFE is purely generated from AO-filtered Kolmogorov turbulence. 
\\\\
Unsurprisingly, we also find that larger lanterns couple more light than smaller lanterns at a given Strehl ratio, because the number of supported modes increases with lantern entrance diameter. However, in the absence of beam-shaping, we show in Fig.~\ref{fig:chrom} that gains from increasing lantern size begin to diminish beyond the 6-port lantern: the absolute coupling gain from a 6- to a 19-port lantern is roughly the same as the gain from a 3- to a 6-port lantern. This result is reiterated by Fig.~\ref{fig:vs_strehl_vs_cd}b, and stems from the fact that low-order modes such as LP$_{01}$, LP$_{11}$, and LP$_{02}$ play a larger role in accepting injected light than higher-order modes, especially when WFE is dominated by low-order aberrations. The LP$_{11}$ modes are well-suited for dealing with tip-tilt aberrations, and account for the increase in coupling efficiency from a SMF to a 3-port lantern. Similarly, in the absence of beam-shaping, the increase in coupling efficiency from a 3 to 6-port lantern is primarily due to the inclusion of the LP$_{02}$ mode (Fig.~\ref{fig:wfs}g), which can accept light from the Airy ring.
\\\\
In the presence of Gaussian beam-shaping optics, diminishing returns begin even sooner: as seen in Fig.~\ref{fig:vs_strehl_vs_cd}a, a 3-port lantern with beam-shaping performs almost as well as a 6-port lantern with or without beam-shaping. As mentioned above, such lenses have a significant effect for three-port lanterns (and SMFs) since such devices have no other way to efficiently accept light from the Airy ring; in contrast, such lenses are less useful for 6-port and larger lanterns, which have access to the LP$_{02}$ mode. However, it is important to recognize that these results are specific to our simple Gaussian-remapping implementation of beam-shaping. Lenses with more advanced remapping functions, tailored to individual lantern mode geometries and imaging systems, may boost coupling more and warrant further exploration. Also of interest are optical systems with two DMs, allowing for simultaneous phase and amplitude modulation without any requirement for radial symmetry. Careful configuration of such setups may boost coupling performance even further.

\subsection{Signal-to-noise considerations}
\begin{figure}
    \centering
    \includegraphics[width=\textwidth]{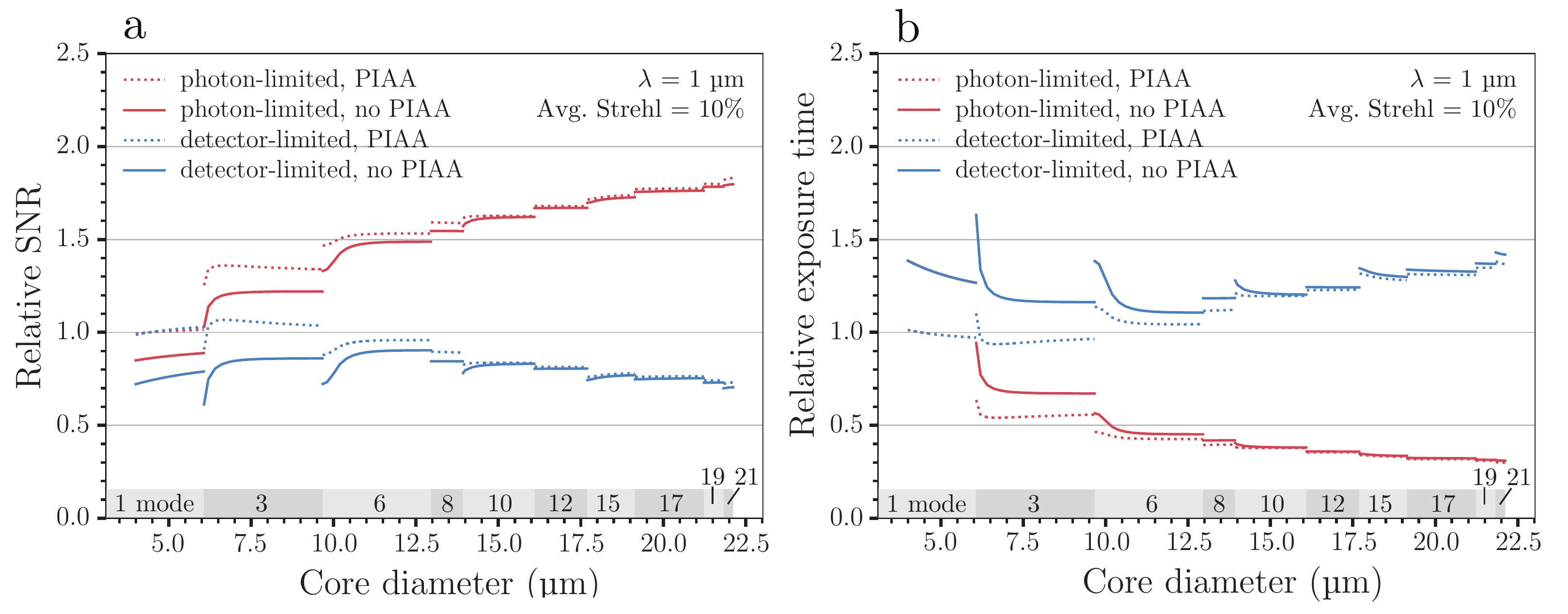}
    \caption{a: SNR of a lantern-fed instrument as a function of entrance diameter at $\lambda = 1$ $\upmu$m, in the photon-limited (red) and detector-limited (blue) noise regimes. Turbulence was fixed to give an average 10\% Strehl. All lantern ports are assumed to be in the same noise regime, and noise is assumed to be spread evenly among the lantern ports. SNR values are relative to the SNR of an SMF with beam-shaping optics. b: the required exposure time for a lantern-fed instrument to attain some reference SNR, relative to the time it would take an SMF with beam-shaping optics to attain that same SNR, as a function of lantern entrance diameter. Wavelength and Strehl are the same as in a. }
    \label{fig:snr}
\end{figure}
While lantern coupling efficiencies are useful for measuring the raw performance of PLs, the signal-to-noise ratio (SNR) is often a more practical metric. Full SNR calculations will depend on instrument specifics and the observing context, so we only consider two limiting cases: the fully photon-limited case, where all lantern ports are dominated by photon noise, and the fully detector-limited case, where all lantern ports are dominated by detector (read, dark current) noise. Figure \ref{fig:snr}a presents the SNR as a function of lantern entrance diameter in these two limits. Atmospheric turbulence is fixed to provide a 10\% Strehl ratio. Here, lantern SNRs are normalized against the SNR of an SMF with beam-shaping optics. In the fully photon-limited (Poissonian) regime, SNR scales as the square root of the lantern coupling efficiency $\mu$: ${\rm SNR}_{\rm photon} \propto \sqrt{\mu}$. In the fully detector-limited regime, the noise variance will be linearly proportional to the number of lantern ports $n_{\rm c}$, penalizing larger lanterns: ${\rm SNR}_{\rm detector}\propto \mu/\sqrt{n_{\rm c}}$. As a result, we see that significant gains in SNR can be made in adopting larger lanterns only if we are in the fully photon-limited regime, with 19-port lanterns reaching an $\sim 80$\% gain over an SMF. Conversely, in the fully detector-limited regime, lantern SNRs are about the same as or worse than that of an SMF. Fig.~\ref{fig:snr}b presents the same argument from the perspective of exposure times. Here, we plot the time required by a lantern-fed instrument to hit some reference SNR, relative to the time it would take an SMF with beam-shaping optics to hit that same SNR. Figure~\ref{fig:snr}b shows that in the photon-limited regime, exposure times for a 3-port lantern with beam-shaping optics are around half that of a similarly configured SMF, and that exposure time halves again going from 3 to 19 ports. In the detector-limited regime, exposure times are increased by a factor of $\sqrt{n_c}$ over the photon-limited regime, so that lanterns generally require more time than an SMF with beam-shaping optics. As such, at 10\% Strehl ratio, if the signal from the lantern ports will be in the detector-limited regime, then an SMF should be used instead. If the output ports will be photon-noise limited, then any lantern can be used and higher port counts will result in greater overall SNR, but only if the signal in each port will remain dominated by photon-noise. In reality, lantern behavior will likely be in between the two limiting cases presented above. If most but not all lantern outputs are in the photon-limited regime, SNR gains from larger lanterns will drop off more quickly than in the fully photon-limited case, strengthening the argument for few-port lanterns. The impact of detector noise may also be lessened with mode-selective lanterns \cite{Leon-Saval:14}, which can direct light so that the bulk of the signal is concentrated within a few high-flux ports, as opposed to being split evenly among many ports.
\\\\
A system level consideration that arises when selecting the number of ports to use is the availability of pixels. For a given set of spectrometer parameters (resolving power and bandwidth), the number of pixels required will increase linearly with the number of SMF ports. For very high resolving power applications ($>$100,000), which typically desire large bandwidths (y-K band for example) as well, port counts using even the largest format NIR detectors will be limited to $\sim$6. Going beyond the footprint of a single detector and requiring a secondary detector or a larger version is typically a costly upgrade that most instruments cannot absorb. In addition, splitting the light across more ports lowers the overall flux per port, requiring longer integration times to regain SNR. As large telescope time is costly, this can also be an undesirable choice. For these reasons, few-port lanterns can be more attractive. Regardless, as Figure \ref{fig:snr} shows, if both pixel constraints and detector noise are not an issue, high mode-count lanterns can provide significant gains in SNR. The cost of introducing adding more ports may also be offset due the increased tip-tilt resilience of larger lanterns. In this case, a large mode-count PL could used instead of high-performance tip-tilt correction, resulting in a trade in system complexity, rather than a strict increase.
\\\\
Finally, it is clear that the beam-shaping optics can work in parallel to the PL to boost coupling. Given that beam-shaping optics can be manufactured for $<$\$5k, such optics can be readily used in conjunction with a lantern to help reduce the required number of pixels. For instance, as seen in Fig.~\ref{fig:chrom}, performance for a 3-port lantern in conjunction with beam-shaping is similar to that of a 6-port lantern with or without beam-shaping optics. By using a 3-port+beam-shaping optics configuration instead of a 6-port lantern, the number of detector pixels can be reduced by a factor of 2! This means that the wavelength coverage could be extended, or large gaps between the traces could be used to minimize cross-talk to improve overall precision. Although the relative cost of the beam-shaping optics is similar to the PL, the addition of this optic can have a tremendous impact compared to the much more expensive detector pixels. 

\subsection{Optimizing coupling efficiency}
After deciding on a targeted number of lantern modes and beam-shaping configuration, the next focus in lantern design is the maximization of input coupling efficiency. This maximization requires the fulfillment of two separate conditions: firstly, the physical extents of the lantern modes and PSF must match; secondly, the beam profile shapes should also match. Condition 1 sets the optimal camera focal ratio (and hence PSF scale) for a given entrance diameter. On the other hand, condition 2 sets the lantern entrance diameter itself. As seen in Fig.~\ref{fig:vs_strehl_vs_cd}b (computed assuming optimal focal ratios for each entrance diameter), over diameter ranges where the number of available modes is constant, coupling efficiency is still variable, implying in turn that the field structure of the lantern modes depends on entrance diameter. This is an analytically known effect, and is most clearly displayed by the fundamental LP$_{01}$ mode, which appears more Lorentzian-like at small entrance diameters, parabolic at large entrance diameters, and Gaussian-like in between. For the same reason, coupling efficiencies can actually drop as core diameter increases. This effect appears to be the strongest in the presence of Gaussian beam-shaping optics (see the 3-mode lantern w/ PIAA in Figure \ref{fig:vs_strehl_vs_cd}b), presumably because increasing core diameter causes mode shapes to deviate further from a Gaussian profile. As such, lantern entrance diameters should be selected to promote lantern modes that are most similar in shape to the PSF. The optimal entrance diameter can be determined via a brute-force search over a range of entrance diameters, like in Fig.~\ref{fig:vs_strehl_vs_cd}b, ensuring that focal ratio is optimized at each tested diameter (otherwise coupling effects due to mismatches in size and shape will be entangled). Ultimately, while effects involving lantern mode shape are not large, they are worth considering during system design and manufacture. 
\\\\
The last parameters to consider in lantern design are the length of the lead-in waveguide and the taper factor, relevant in regimes where the number of lantern ports strictly exceeds the number of propagating modes at the lantern's entrance (e.g. Fig.~\ref{fig:leadin} for wavelengths beyond 1.3 $\upmu$m). In such regimes, weakly attenuating radiative modes can still help transfer light to the lantern cores, promoting lantern throughputs that exceed the coupling efficiency into the guided modes at the lantern's entrance. A similar experimental effect is mentioned in Birks et al. \cite{Birks:15}, which notes that prior lantern experiments have found throughputs that were better than expected in wavelength regimes where lantern operation was no longer strictly efficient. As a potential explanation Birks et al. \cite{Birks:15} proposed that experimental setups may not have excited all available modes in their lanterns; we offer the alternative explanation that instead the presence of weakly radiative modes at a lantern's entrance may be boosting the effective modality of the lanterns. In turn, lantern designs in wavelength regimes near and beyond transition wavelengths should carefully consider lead-in length and taper factor: the tuning of these parameters presents an avenue through which the bandwidth for efficient lantern operation can be maximized. Optimal performance occurs when there is no lead-in, corresponding to direct injection of light into the lantern's transition zone: this approach minimizes the attenuation of any light travelling in the radiative modes. From a manufacturing perspective, our result implies that lanterns should be cleaved as close to the transition zone as possible. For the particular case of the 3-port lantern in Fig.~ \ref{fig:leadin}, we have shown that such a lantern can maintain its modality over the entirety of the $y$ and $J$ bands if the lead-in waveguide length is less than 0.01 cm. 
\\\\
As seen in Fig.~\ref{fig:atten}, attenuation of the radiative lantern modes can also be reduced by lowering the lantern taper factor, which in turn increases the size of the embedded SMF cores at the lantern entrance and hence lowers the imaginary effective refractive index at that location. In practice, obtaining smaller taper factors with SMF-based lanterns often involves etching down the claddings of SMFs or pre-tapering the individual SMFs before stacking and tapering the bundle. Otherwise, taper factors will tend to be quite large: for instance, a 3-core lantern constructed from a bundle of 3 SMFs with cladding diameters of 125 $\upmu$m requires a taper factor of $\sim$34 to yield a lantern entrance diameter of 8 $\upmu$m. Taper factors can also be lowered somewhat by instead constructing lanterns from MCFs. For reference, commercially available 4-core MCFs such as Chiral Photonics MCF-004\_1 or Fibercore SM-4C1500 have typical cladding diameters of $\sim 125$ $\upmu$m, requiring only a $\sim$16$\times$ taper to produce an 8 $\upmu$m entrance diameter. The required taper factors to construct a 3-port lantern from a 3-core MCF will likely be similar. Beyond radiative attenuation, taper factor will also affect lantern throughput in another more indirect way: the presence of large residual SMF cores in the lead-in waveguide (corresponding to a small taper factor) will change the shape of the lantern modes to diverge from that of the LP modes, impacting the initial coupling efficiency into the lantern. We do not treat this higher-order effect, and leave it for future work. 
\\\\
Finally, radiative mode attenuation may also be reduced by shortening a lantern's transition zone, though this effect is minor since light only has to travel through a fraction of the transition zone before the lantern cladding becomes large enough to support the previously radiative LP modes. However, care must also be taken not to make the lantern too short, otherwise the transformation of light from input to output will not be adiabatic, and performance will drop \cite{Birks:15}. The tension between adiabaticity and radiative mode attenuation implies that lanterns have an optimal taper length (see \cite{Corrigan:18}), which should be solved for during lantern design.
\subsection{Future directions}
The next step will be to verify the results of our simulations on-sky. While we have conducted an initial investigation on the interaction between tip-tilt errors and PLs, real wavefronts have extra low-order error components that are not captured by our simulations. On-sky verification will also allow us to extend our results beyond the idealized, linearly tapered and perfectly circular lanterns assumed in this work. While we believe our simulations accurately capture the first-order behavior of PLs, these higher-order effects will need to be measured before PLs can be fully integrated into the next generation of instruments.  

\section{Conclusion}
In this work we have presented numerical simulations characterizing the potential performance gains of lantern-based fiber injectors and their interaction with beam-shaping optics over a range of Strehl ratios, wavelengths, and lantern geometries. Lantern performance was first characterized in terms of coupling efficiency into the lantern input. This proxy metric sets the first-order behavior of the PL and serves as a good approximation for overall lantern throughput over wavelengths where the number of guided modes at the lantern's entrance remains constant. We find that PLs show a linear scaling in coupling efficiency with Strehl ratio when WFE is solely generated from Kolmogorov turbulence. At low Strehl ($\sim$10\%), and in the absence of beam-shaping, coupling efficiency into the simplest 3-port (8 $\upmu$m entrance diameter) lantern is around twice that of an SMF. A 6-port lantern (12.4 $\upmu$m entrance diameter) and 19-port lantern (21.8 $\upmu$m) triple and quadruple the performance of an SMF, respectively. When lanterns are combined with Gaussian beam-shaping (PIAA) optics, we find that the relative boost offered by such optics diminishes rapidly, such that 3-port lanterns and 6-port lanterns now perform similarly at low Strehl. This drop-off occurs because lanterns with 6 modes or more have access to the LP$_{02}$ mode, which can accept light from the Airy ring, in turn lessening the utility of beam-shaping. At high Strehl ($\sim 70$\%), in the presence of beam-shaping, lanterns and SMFs perform similarly. We also show that PLs are resilient to tip-tilt aberrations, scaling with the square root of Strehl when WFE is modulated via injected tip-tilt. As such, in real optical systems, where WFE has additional low-order components from NCPAs and mechanical vibrations, the gains from PLs may be even greater. From our analysis of lantern coupling efficiencies, we find that for operation in the astronomical $y$ and $J$ bands when Strehl is low ($\sim$ 10\%), 3-port lanterns in conjunction with beam-shaping strike a good balance between performance and pixel count, offering nearly the same coupling efficiency as a 6-port configuration with half the lantern outputs. If pixels are not a constraint, 19-port lanterns can provide even greater coupling efficiencies and better resilience to tip-tilt jitter, doubling that of a 3-port lantern with beam-shaping optics at $\lambda = 1$ $\upmu$m, as long as all ports have sufficient flux to be dominated by photon noise.
\\\\
Over wavelengths where the number of guided modes at the lantern entrance drops, we find that the performance of PLs may not necessarily suffer. In fact, past such wavelengths, lanterns with sufficiently short lead-in waveguide lengths will act as to retain their modality due to the presence of weakly radiative modes at the lantern entrance. To promote this behavior and ensure optimal lantern performance across large wavelength bands where the number of guided modes at the lantern entrance changes, we find that lantern lead-in lengths should be minimized in order to keep radiative losses low. Maximal performance occurs when telescope light is directly injected into the transition zone of the lantern, and when lantern taper factors are small. Future instruments that adopt PLs should carefully consider these geometric factors to ensure optimal performance over the widest range of wavelengths.

\section*{Funding}
The authors acknowledge the support of the Heising-Simons Foundation (award \#2020-1821).
 
\section*{Acknowledgments}

The authors thank G. Ruane for providing the PIAA lens design. The authors would also like to thank C. Betters, S. Leon-Saval, and B. Norris for valuable discussions on photonic lanterns, as well as O. Guyon for his insight on various components of the numerical model.

\section*{Disclosures}

The authors declare no conflicts of interest.

\bibliography{literature}

\end{document}